\documentclass[aps,prl,amsmath,reprint,superscriptaddress]{revtex4-2}
\usepackage{graphicx}
\usepackage{dcolumn}
\usepackage{bm}
\usepackage[colorlinks=true, linkcolor=blue, citecolor=blue, urlcolor=blue,breaklinks=true]{hyperref}
\usepackage{nameref}
\usepackage[nameinlink]{cleveref}
\usepackage[mathlines]{lineno}
\usepackage{amsthm,amssymb}
\usepackage{physics}
\usepackage{mathrsfs}
\usepackage{mathtools}
\usepackage{xcolor}
\usepackage[caption=false]{subfig}
\usepackage{soul}
\usepackage{colortbl}
\usepackage[vlined,boxed]{algorithm2e}

\makeatletter
\def\@fnsymbol#1{\ensuremath{\ifcase#1\or \dagger\or \ddagger\or
\mathsection\or \mathparagraph\or \|\or **\or \dagger\dagger
\or \ddagger\ddagger \else\@ctrerr\fi}}
\makeatother

\footnotetext{These authors contributed equally.}

\rmfamily

\definecolor{bur}{rgb}{0.5, 0.0, 0.13}
\definecolor{purple}{RGB}{128,0,128}

\newcommand\blk[1]{\color{black}#1}

\newcommand{\parTitle}[1]{
\emph{#1}.---}

\begin{document}
\title{Quantum assemblage tomography}

\author{Luis Villegas-Aguilar$^*$}
\email{luis@villegasaguilar.com}
\affiliation{Centre for Quantum Dynamics and Centre for Quantum Computation and Communication Technology (CQC$\,^2\!$T), Griffith University, Yuggera Country, Brisbane, 4111 Australia}

\author{Yuanlong Wang$^*$}
\email{wangyuanlong@amss.ac.cn}
\affiliation{
State Key Laboratory of Mathematical Sciences, Academy of Mathematics and Systems Science, Chinese Academy of Sciences, Beijing 100190, People’s Republic of China}
\affiliation{School of Mathematical Sciences, University of Chinese Academy of Sciences, Beijing 100049, People’s Republic of China}
\affiliation{Centre for Quantum Dynamics and Centre for Quantum Computation and Communication Technology (CQC$\,^2\!$T), Griffith University, Yuggera Country, Brisbane, 4111 Australia}

\author{Alex Pepper$^*$}
\affiliation{Centre for Quantum Dynamics and Centre for Quantum Computation and Communication Technology (CQC$\,^2\!$T), Griffith University, Yuggera Country, Brisbane, 4111 Australia}

\author{Travis J. Baker}
\affiliation{Centre for Quantum Dynamics, Griffith University, Yuggera Country, Brisbane, 4111 Australia}
\affiliation{%
Nanyang Quantum Hub, School of Physical and Mathematical Sciences, Nanyang Technological
University, Singapore 637371
}%
\author{Dominick J. Joch}
\affiliation{Centre for Quantum Dynamics and Centre for Quantum Computation and Communication Technology (CQC$\,^2\!$T), Griffith University, Yuggera Country, Brisbane, 4111 Australia}
\author{Sven Rogge}
\affiliation{Centre for Quantum Computation and Communication Technology (CQC$\,^2\!$T), School of Physics, The University of New South Wales, Sydney, NSW 2052, Australia}
\author{Geoff J. Pryde}
\author{Sergei Slussarenko}
\author{Nora Tischler}
\author{Howard M. Wiseman}
\affiliation{Centre for Quantum Dynamics and Centre for Quantum Computation and Communication Technology (CQC$\,^2\!$T), Griffith University, Yuggera Country, Brisbane, 4111 Australia}

\date{\today}

\begin{abstract}
A central requirement in asymmetric quantum nonlocality protocols, such as quantum steering, is the precise reconstruction of state assemblages---statistical ensembles of quantum states correlated with remote classical signals.
{Existing steering works often rely on simplifying assumptions about detection efficiency and photon loss.}
Here we introduce a generalized loss model for assemblage tomography that uses conical optimization techniques combined with maximum likelihood estimation.
{This approach allows us to accurately estimate assemblages without assuming uniform detection efficiency on the untrusted party's side.
Using an evidence-based framework grounded in the Akaike information criterion, we demonstrate faithful reconstructions while balancing model complexity.
We validate our results through numerical simulations and an experimental setup, showing robust performance in assemblage estimation when applied to experimentally relevant data.}
\end{abstract}
\maketitle

\parTitle{Introduction}Recent advances in quantum technologies have proven the ability to meticulously engineer and control quantum systems, offering potential advantages in areas like computation, communication, and metrology~\cite{acin2018QuantumTechnologiesRoadmap}.
As quantum technologies continue to mature, a crucial challenge arises in the reliable characterization of diverse quantum resources~\cite{eisert2020QuantumCertificationBenchmarking}.
A standard strategy for this task involves using tomographic techniques \cite{vogel1989DeterminationQuasiprobabilityDistributions, james2001MeasurementQubits}, which allow for reconstructing unknown quantum systems from finite measurement data.
Despite substantial research focusing on estimation techniques for quantum states \cite{dariano2001QuantumTomographyMeasuring, cramer2010EfficientQuantumState, gross2010QuantumStateTomography, torlai2018NeuralnetworkQuantumState} and processes \cite{chuang1997PrescriptionExperimentalDetermination, obrien2004QuantumProcessTomography, mohseni2008QuantumprocessTomographyResource}, there remains a lack of robust methods for the tomography of quantum state assemblages, the central resource used to define quantum steering \cite{cavalcanti2016QuantumSteeringReview, uola2020QuantumSteering}.

Quantum (or Einstein-Podolsky-Rosen) steering~\cite{jones2007EntanglementEinsteinPodolskyRosenCorrelations} captures the ability to remotely influence, or \textit{steer}, the state of a quantum system by performing local measurements on a second, distant system.
From a quantum information perspective, quantum steering offers promise for scalable and resource-efficient 
secure communication protocols~\cite{bennet2012ArbitrarilyLossTolerantEinsteinPodolskyRosen, kocsis2015ExperimentalMeasurementdeviceindependentVerification, cavalcanti2015DetectionEntanglementAsymmetric, li2015CertifyingSinglesystemSteering}.
Quantum steering can be demonstrated from quantum state assemblages---ensembles of locally accessible states conditionally prepared by a remote party. \blk 
A standard quantum steering scenario involves the joint outcomes of an uncharacterized and potentially inaccessible measurement station and a trusted device performing known quantum measurements.
Using the empirical frequencies derived from this joint dataset, one can optimally estimate an assemblage in the sense of minimizing a suitable cost function on the space of assemblages. 
Once an assemblage is estimated, it becomes possible to demonstrate the presence of steering through semidefinite programming~\cite{Pusey13, cavalcanti2016QuantumSteeringReview}, to directly analyze correlations through quantum steering ellipsoid geometry~\cite{Jevtic14, Cheng16, Xu24}, or to calculate other assemblages accessible by physically allowed transformations in the context of resource theories~\cite{Gal15, Zjawin2023quantifyingepr}. 

Unlike quantum state tomography, which typically assumes complete control over all measurement devices, a fully rigorous approach to assemblage tomography must necessarily account for losses associated with the uncharacterized devices to ensure faithful reconstructions.

Here, we present a generalized approach to quantum assemblage estimation via a maximum-likelihood estimation (MLE) \cite{hradil2004MaximumLikelihoodMethodsQuantum} seesaw algorithm, using conical optimization techniques~\cite{boyd}; our code is openly accessible~\cite{github}.
We address the realistic scenario where uncharacterized measurement devices may occasionally produce inconclusive outcomes, e.g.,
{
because of particle loss, inefficient state readout~\cite{antoniadis2023CavityenhancedSingleshotReadout, gritsch2025OpticalSingleshotReadout}, finite system lifetimes~\cite{lambrecht2017LongLifetimesEffective}, or leakage errors~\cite{miao2023OvercomingLeakageQuantum}.}
Due to the inherently asymmetric nature of steering tasks, it is crucial to treat such no-detection events as an additional outcome of the measurement process~\cite{czechlewski2018InfluenceChoicePostprocessing}, therefore treating ``no-click" outcomes on par with conclusive detection events to faithfully estimate state assemblages.

We introduce a general loss model where detection efficiencies in uncharacterized devices need not be uniform for different measurement settings or between different detectors, providing a more accurate representation of experimental conditions.
By adopting a model selection framework~\cite{akaike1998InformationTheoryExtension, burnham2004ModelSelectionMultimodel}, we are able to answer the question ``to what extent is the estimated assemblage consistent with the observed experimental data?''~with a quantified measure of statistical evidence. 
We numerically show that our comprehensive loss model yields assemblages that accurately reflect the empirical data, effectively minimizing bias when estimating their steerability.

\parTitle{Preliminaries}\label{sec:preliminaries}Suppose we have a scenario in which an uncharacterized source prepares a sequence of bipartite quantum systems, each in the state $\rho_{\rm AB}$.
In each round, one copy of $\rho_{\rm AB}$ is distributed between distant measurement stations, Alice and Bob, with Bob's station being trusted.
Alice performs a measurement determined by a classical variable $x$, where $x\in\{0,1,\ldots,m-1\}$. She obtains an outcome $a$, with probability $p_{a|x}$, and communicates this outcome to Bob. The state of Bob's system, conditioned on the observation of said outcome $a$, is $\rho_{a|x}$. 
Over many rounds, Bob's goal is to estimate the $m$ different ensembles, $\{ \sigma_{a|x} \}_{a,x}$, 
where $\sigma_{a|x} = p_{a|x}\rho_{a|x}$, by using a well-characterized measurement device to probe his local system.
The collection of these $m$ ensembles is referred to as an \textit{assemblage}.
Each unnormalized state $\sigma_{a|x}$ is given by 
\begin{equation}
\label{eqn:obtaining_assemblages}
  \sigma_{a\vert x} = \text{Tr}_{\rm A} \left[ \left(E_{a\vert x}\otimes \mathbb{I}_{\rm B}\right)\rho_{\rm AB}\right],
\end{equation}
where, for each $x$, $\{E_{a|x}\}_a$ is Alice's (uncharacterized) positive operator-valued measure (POVM) over $a$, satisfying $\sum_a E_{a\vert x}=\mathbb{I}_A$ and $E_{a\vert x}\geq 0, \ \forall a, x$.
 
Given that Alice and Bob are assumed to conduct measurements on spatially separated systems, any assemblage prepared for Bob must satisfy no-signaling (NS) constraints. This property is formalized by the relations
\begin{equation}
\label{eq:no_signaling_cond}
 \sum_a \sigma_{a\vert x}  = \rho_{\rm B} \ \ \ \forall x,
\end{equation}
where $\rho_{\rm B} \coloneqq \text{Tr}_{\rm A}(\rho_{\rm AB})$.
This ensures that Bob's marginal state cannot be influenced by Alice's choice of measurement setting.
The naive approach to assemblage tomography---individually reconstructing each $\sigma_{a\vert x}$ from the observed data---will almost certainly fail to satisfy Eq.~\eqref{eq:no_signaling_cond} due to experimental noise and finite statistics.
This motivates the development of assemblage tomography techniques that can yield valid, no-signaling assemblages capable of reproducing the observed data as closely as possible.

\bgroup
\def\arraystretch{1.5}
\begin{table*}
\begin{tabular}{c c c c c c c}
Model & \ & Efficiency & \ & Bias & \ & Unknown parameters $\boldsymbol{\theta}$ \\
\hline
M1 &  \ \ \  & $\epsilon_x = \epsilon = \text{const.}, \ \forall x$ &  \ \ \  & $\gamma_x = 0, \ \forall x$ &  \ \ \  & $\rho_{\rm B}, \ \{\sigma_{a\vert x}\}_{a,x}$ \\
M2 &  \ \ \  & $\boldsymbol\epsilon = \left(\epsilon_0,\ldots, \epsilon_x, \ldots, \epsilon_{m-1}\right)$ &  \ \ \  & $\gamma_x = 0, \ \forall x$ &  \ \ \  & $\rho_{\rm B}, \ \{\sigma_{a\vert x}\}_{a,x}, \ \boldsymbol \epsilon$ \\
M3 &  \ \ \  & $\boldsymbol\epsilon = \left(\epsilon_0,\ldots, \epsilon_x, \ldots, \epsilon_{m-1}\right)$ &  \ \ \  & $\boldsymbol\gamma = \left(\gamma_0,\ldots, \gamma_x, \ldots, \gamma_{m-1}\right)$ & \ \ \ & $\rho_{\rm B}, \ \{\sigma_{a\vert x}\}_{a,x}, \ \boldsymbol \epsilon, \ \boldsymbol \gamma$ \\
\end{tabular}
\caption{Maximum-likelihood assemblage tomography algorithms under different loss models.
These models can range from a single, static efficiency $\epsilon$ (M1) to a more sophisticated model (M3) where the detection efficiency and bias depend on the measurement setting ($\epsilon_x, \gamma_x$).
}
\label{tab:table1}
\end{table*}
\egroup

\parTitle{Dealing with nondetection events}In practical experimental implementations of the scenario described above, Alice's detectors can also produce nondetection, or null, events due to losses in the transmission or measurement processes. 
Null outcomes could also stem from a dishonest Alice attempting to skew the joint statistics by selectively reporting outcomes only in certain measurement rounds.
One approach to handling this issue is to discard measurement rounds where Alice does not obtain one of the outcomes $a$. 
Doing this, however, amounts to Bob invoking a \emph{fair sampling assumption}~\cite{pearle1970HiddenVariableExampleBased} about how the measurement outcomes are generated.
This introduces loopholes into nonlocality tests~\cite{pearle1970HiddenVariableExampleBased, santos1992CriticalAnalysisEmpirical, larsson2014LoopholesBellInequality} and compromises the security of cryptographic applications that rely on quantum steering correlations certified by the assemblage formalism~\cite{uola2020QuantumSteering}. 
Instead, it is now a standard technique to account for nondetection events by either assigning them the same label as one of the conclusive detection outcomes~\cite{Evans13, sun2018DemonstrationEinsteinPodolsky, dilley2018MoreNonlocalityLess} or to treat them as an additional outcome of Alice's measurements~\cite{wittmann2012LoopholefreeEinsteinPodolsky, zeng2022OnewayEinsteinPodolskyRosenSteering}.
We opt for the latter approach.
We consider the case where Alice performs dichotomic measurements on a system that may be lost, so that her measurements yield outcome variables $a\in\{+,-,\emptyset\}$, with $\emptyset$ indicating a null event reported by Alice. 

A crucial challenge is determining how losses in Alice's measurements affect Bob's ability to reconstruct the assemblage while ensuring that the NS conditions are satisfied.
To model this, we assume that Alice's system could be lost at \emph{any} stage before her detectors register an outcome.
Suppose, for instance, that her system is lost during its interaction with her device set to measure $x$.
{
This loss can be described by transforming the initial state of her system as $\rho_{\rm A} \rightarrow \tilde{\rho}_{\rm A} \coloneqq \epsilon_{x} \rho_{\rm A} + (1-\epsilon_{x}) \ketbra{v}$, where $\epsilon_x$ is the efficiency of her measurement device for setting $x$ and $\ket{v}$ denotes a state orthogonal to the support of $\rho_\mathrm{A}$, representing the ``lost'' component of the system.
Moreover, the detectors themselves may have inefficiencies that cause variations in the observed frequencies of the $+$ and $-$ outcomes for the same measurement (i.e., Alice's detectors may be unbalanced).
These effects are captured by bias parameters $\gamma_x \in [-(1-\epsilon_x), 1-\epsilon_x]$, which account for variations in the probabilities of non-null outcomes as
\begin{equation}
\label{eq:gamma_f}
p_{\pm|x} = \left[\epsilon_x + f_\pm(\gamma_x)\right] \mathrm{Tr}[E_{\pm|x}\rho_\mathrm{A}].
\end{equation}
Here, $\epsilon_x$ can be interpreted as a ``baseline efficiency'' for setting $x$, and the functions $f_\pm(\gamma_x) := (|\gamma_x| \pm \gamma_x)/2$, which equal either $\gamma_x$ or $0$, model imbalance in the statistics of non-null outcomes.
This is to account for potential differences in the detection efficiencies of Alice's measurement.
While alternative parametrizations are possible, any approach that accounts for setting- and outcome-dependent differences must be mathematically equivalent to the above.
}
The general model for losses we have described thus far can also be viewed in the Heisenberg picture, where $\text{Tr}(E_{a|x}\tilde{\rho}_{\rm A})\equiv\text{Tr}(\tilde{E}_{a|x}\rho_{\rm A})$.
Here, each of Alice's measurements now corresponds to $\tilde{E}_{\pm|x} \coloneqq \left[\epsilon_x + f_\pm (\gamma_x)\right]E_{\pm|x}$ and  $\tilde{E}_{\emptyset|x} = \mathbb{I}_{\rm A} - \sum_{a\neq\emptyset} \tilde{E}_{a|x}$ being performed on her original system. Consequently, the elements of the assemblage prepared for Bob will inherit the following structure, where, for $a\in\{+,-\}$,
\begin{equation}
\Tr_{\rm A} \left[ \left(\tilde{E}_{\pm|x} \otimes \mathbb{I}_{\rm B}\right) \rho_{\rm AB} \right] = 
\left[\epsilon_x + f_\pm (\gamma_x)\right] \sigma_{\pm|x},
\label{eq:steered_non_nulls}
\end{equation}
and, for the null result,
\begin{equation}
\sigma_{\emptyset|x}=(1-\epsilon_x)\rho_{\rm B} + f_+(\gamma_x)\sigma_{+|x} + f_-(\gamma_x) \sigma_{-|x}.
\label{eq:steered_nulls}
\end{equation}

\parTitle{Maximum-likelihood estimation}We formulate Bob's tomographic task as estimating the most probable assemblage prepared for him, while respecting NS and considering the impacts of loss on Alice's measurements according to Eqs.~\eqref{eq:steered_non_nulls} and \eqref{eq:steered_nulls}.
To be precise, we treat all the efficiencies $\boldsymbol \epsilon = (\epsilon_0,\ldots,\epsilon_x,\ldots, \epsilon_{m-1})$ and biases $\boldsymbol \gamma = (\gamma_0,\ldots,\gamma_x,\ldots, \gamma_{m-1})$ as additional unknowns to be estimated in the assemblage tomography process.

The general tomography problem can then be framed as finding the most likely assemblage $\{\sigma_{a|x}\}_{a,x}$, reduced state $\rho_{\rm B}$, and parameters $\boldsymbol \epsilon$ and $\boldsymbol\gamma$, satisfying Eq.~\eqref{eq:steered_nulls} and the following constraints:
\begin{align}
\label{equ1}
\sigma_{a|x}&\geq 0 \ \ \forall x,a,  \\
\label{equ2}
\sum_a\sigma_{a|x}&=\rho_{\rm B} \ \ \forall x,\\
\label{equ4}
\text{Tr}(\rho_{\rm B})&=1, \\
\label{equ5}
|\gamma_x|&\leq  1 - \epsilon_x \ \ \forall x. 
\end{align}

Given the above, the logarithmic likelihood for any candidate assemblage to generate the given experimental data will be 
\begin{equation}\label{eqMLE}
  \log[\mathcal{L}] = \sum_{a,b,x,y}N(b|y)|_{a|x}\log\left[\text{Tr}(E_{b|y}\sigma_{a|x})\right],
\end{equation}
where $N(b|y)|_{a|x}$ is the experimental count corresponding to the POVM $E_{b\vert y}$ performed on Bob's side, conditioned on $x$ and $a$.

Our main result corresponds to a comprehensive assemblage tomography model, referred to as model 3 (M3), for scenarios where Alice's detection events adhere to the loss scenario outlined so far. Starting from M3, we also consider two simpler, nested models that vary in complexity based on the assumptions made about loss in the system. Tab.~\ref{tab:table1} summarizes our approach.
If $\gamma_{x} = \gamma = 0$, the loss model remains setting-dependent but does not account for imbalances in detection outcomes. This is model 2 (M2), which aligns with assumptions made in recent steering works (see Refs.~\cite{joch2022CertifiedRandomnumberGeneration, zeng2022OnewayEinsteinPodolskyRosenSteering, pepper2024ScalableMultipartySteering}).
A further simplification occurs when $\epsilon_{x} = \epsilon$, resulting in a model with a single global detection efficiency~\cite{bennet2012ArbitrarilyLossTolerantEinsteinPodolskyRosen, smith2012ConclusiveQuantumSteering}; this is denoted as model 1 (M1).

For each model $k$, the optimization of $\log\left[\mathcal{L}\right]$ is performed over a vector of unknown parameters $\boldsymbol{\theta}_k = (\theta_0, \theta_1, \ldots, \theta_{p_k})$ as per Tab.~\ref{tab:table1}.
Although $-\log[\mathcal{L}]$ is convex, optimizing it is a very difficult task because the constraint in Eq.~\eqref{eq:steered_nulls} is nonlinear (it is, in fact, nonconvex).
{
We address this problem by using a two-layer iterative procedure.
In the ``outer-layer'', we optimize $\log[\mathcal{L}]$ over $(\rho_{\rm B}, \{\sigma_{a\vert x}\}_{a,x})$, given fixed values for $\boldsymbol{\epsilon}$ and $\boldsymbol{\gamma}$.
A second, ``inner-layer'' loop uses the estimated $\rho_\mathrm{B}$ to independently optimize $(\boldsymbol \epsilon, \boldsymbol\gamma, \{\sigma_{a\vert x}\}_{a,x})$.
Both layers use conic optimization to estimate a subset of the total unknown parameters, and the process iteratively alternates between the two layers until overall convergence is achieved.
}
A description of the full tomography algorithm is in the Supplemental Material (SM)~\cite{SI} (also see~\cite{peters2004MixedstateSensitivitySeveral, nery2020DistillationQuantumSteering, schumacher1996SendingEntanglementNoisy, oreshkov2009DistinguishabilityMeasuresEnsembles, tavakoli2024QuantumSteeringImprecise, hurvich1989RegressionTimeSeries, tischler2018ConclusiveExperimentalDemonstrationa, jozsa1994FidelityMixedQuantum, esmaeilzadeh2017SinglephotonDetectorsCombininga}) 
and our code is freely available~\cite{github}.
{Optimizing Eq.~\eqref{eqMLE} results in point estimates of the vector $\boldsymbol{\theta}_k$, but one can readily adapt techniques used in quantum state tomography~\cite{blume-kohout2012RobustErrorBarsa, christandl2012ReliableQuantumState, wang2019ConfidencePolytopesQuantum} to instead derive region estimates for the objectives of assemblage tomography.}

\parTitle{Optimal model selection}Any statistical model is necessarily a simplification of physical reality and cannot capture every detail. Therefore, our goal is to identify an assemblage tomography model that well approximates the experimental data while mitigating the possibility of overfitting~\cite{lever2016ModelSelectionOverfitting}. Thus, we are guided by the principle of parsimony in which the best model should be made as simple as it can be, but not simpler.

In the language of information theory, encoding a model involves an inevitable loss of information about the underlying physical process, which is measured by the relative entropy~\cite{kullback1951InformationSufficiency} between the true probability distribution $P$ (that generates the experimental dataset $\mathbb{D}$) and a statistical model approximating $P$.
Since $P$ is unknown, the exact relative entropy cannot be calculated directly.
However, for sufficiently large data sets (see SM~\cite[Note~IV]{SI}), we can estimate the \textit{relative} information loss between competing models and $P$ using the Akaike information criterion (AIC)~\cite{akaike1998InformationTheoryExtension, burnham2004ModelSelectionMultimodel}.
The AIC has been used for quantum state estimation~\cite{usami2003AccuracyQuantumstateEstimation, yin2011InformationCriteriaEfficient, guta2012RankbasedModelSelection, schwarz2013ErrorModelsQuantum, enk2013WhenQuantumTomography, yano2023QuantumInformationCriteria}, and is defined as $\mathrm{AIC} = -2\log\left[\mathcal{L}(\hat{\boldsymbol{\theta}})\right]+2p$, where $\hat{\boldsymbol{\theta}}$ is obtained via the MLE of Eq.~\eqref{eqMLE} and $p$ is the number of free parameters in the model.
{The dominant contribution to $p$ comes from the cardinality of the assemblage ($9m$), with additional parameters accounting for the efficiency ($m$, for M2 and M3) and bias ($m$, for M3) vectors, as per Tab.~\ref{tab:table1}.}

The AIC balances goodness-of-fit and model complexity, with the model having the smallest AIC considered the most likely to have produced $\mathbb{D}$.
{Information criteria like the AIC and other formulations, including the Bayesian information criterion~\cite{neath2012BayesianInformationCriterion, rehacek2008TomographyQuantumDiagnostics}, are powerful statistical tools for ranking competing tomography models.
In experiments, the ground truth (i.e., the true underlying model behind $\mathbb{D}$) is typically inaccessible.
This provides a clear motivation for evidence-based approaches, despite being heuristics that cannot universally guarantee optimality~\cite{ding2018ModelSelectionTechniques}.
Alternatives like fidelity-based benchmarks require a known ground truth for comparison (SM~\cite[Note~II]{SI}).
}

To rank $k$ different tomography models, we compute the differences
  \begin{equation}
  \label{eq:aic}
  \Delta \mathrm{AIC}_k = \mathrm{AIC}_k - \mathrm{AIC}_\mathrm{min},
  \end{equation}
where $\mathrm{AIC}_\mathrm{min}$ is the minimum AIC among competing models for the same dataset $\mathbb{D}$.
The relative likelihood between model $k$ and the best model is $\mathrm{exp}(-\Delta\mathrm{AIC}_k/2)$. Consequently, models become exponentially less supported by $\mathbb{D}$ as $\Delta\mathrm{AIC}_k$ increases~\cite{burnham2011AICModelSelection}.
By calculating $\Delta\mathrm{AIC}_k$ for the various tomographic models under different experimental conditions, we can identify which model is best supported by the experimental data.

\begin{figure}
  \centering
  \includegraphics[width=7cm]{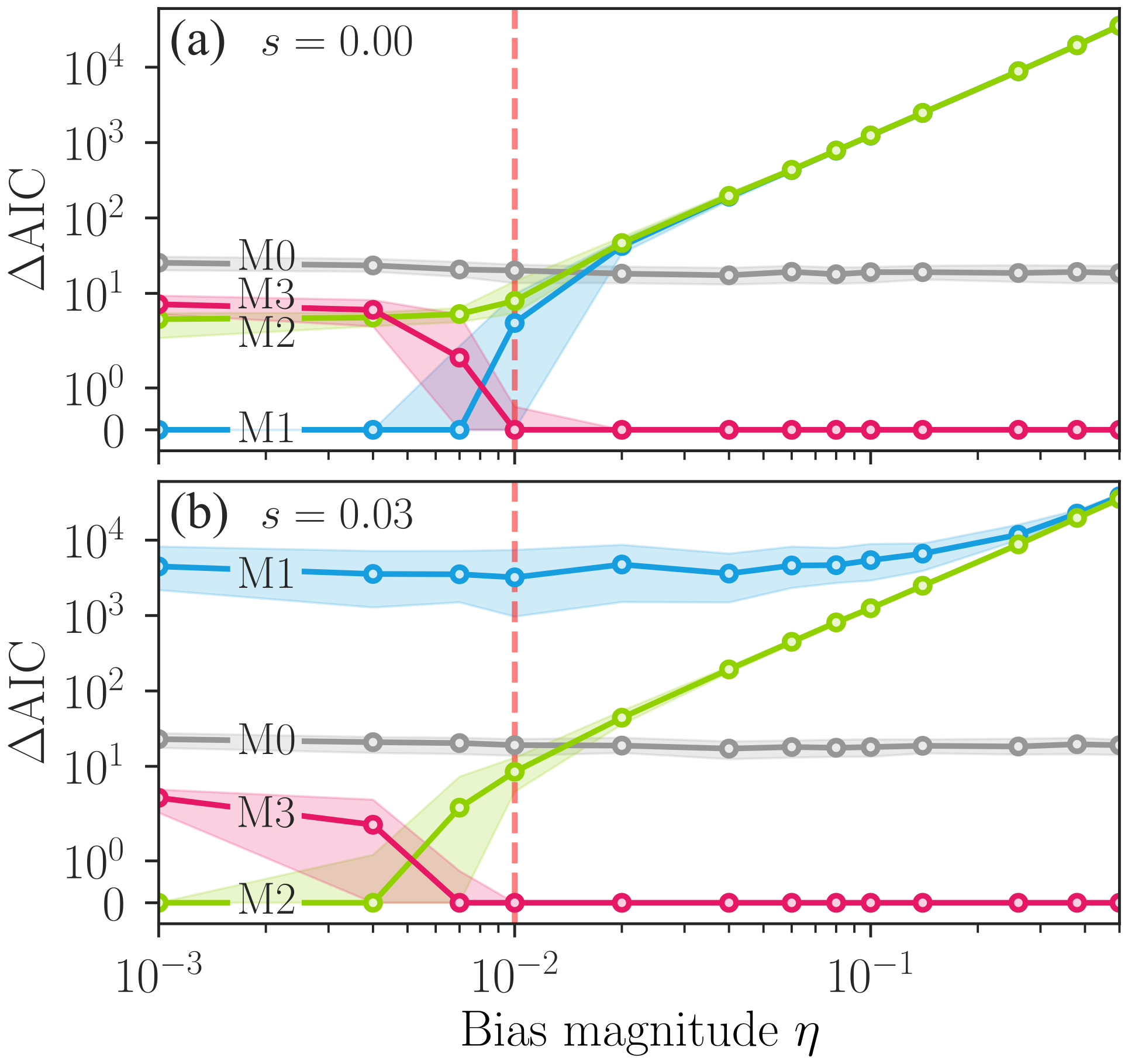}
  \caption{
  Comparison between assemblage tomography models under realistic experimental conditions.
  Numerical simulations depict $\Delta\mathrm{AIC}$ vs detector bias magnitude $\eta$ for zero (a) and nonzero (b) setting-dependent effects.
  Results from 500 Monte Carlo simulations are aggregated and shown as median (solid lines) with interquartile ranges (shaded regions).
  Lower $\Delta\mathrm{AIC}$ values indicate stronger statistical support for a model.
  The vertical dashed line marks the $1\%$ detector bias magnitude threshold.}
  \label{fig:werner_results}
\end{figure}
\parTitle{Numerical studies and experimental results}We analyze the impact of detector bias in assemblage estimation through numerical simulations and with experimental data. \blk
We focus on the assemblages prepared for Bob when Alice measures the Pauli observables $X$, $Y$, and $Z$ in the case of systematic bias with $\gamma_x \geq 0 \ \forall x$ (see SM~\cite[Note~III]{SI}).
{
We first introduce a dependence on Alice's measurement setting by assuming that, for each $x$, the values of $\epsilon_x$ are normally distributed around a mean value $\epsilon$.
To incorporate systematic bias, we modify the efficiency of one detector per setting by adding a factor proportional to $(1-\epsilon_x)$, reflecting the fact that the offset functions $f_\pm$, defined following Eq.~\eqref{eq:gamma_f}, are nonnegative.
}

This effect is parametrized by
\begin{equation}
\label{eq:parametrization}
\gamma_x = \eta(1-\epsilon_x) \ \forall x,
\quad \text{with} \ \
\epsilon_x \sim \mathrm{Norm}_{5}(\epsilon, s),
\end{equation}
where $s$ quantifies the spread of Alice's setting efficiency around $\epsilon$ (with $\mathrm{Norm}_{5}$ denoting a truncated Gaussian distribution within $[-5s, 5s]$), and
$\eta \in [0,1]$ controls the overall magnitude of her detector bias.
{
As a result, Alice's “effective detection efficiency” for positive outcomes becomes $\epsilon_x\rightarrow\epsilon_x + \eta(1-\epsilon_x)$, while the efficiency for negative outcomes remains unchanged.
By varying $s$ and $\eta$, one recovers the conditions corresponding to M1 ($s=0$, $\eta=0$), M2 ($s\neq0$, $\eta=0$), and M3 ($s\neq0$, $\eta\neq0$).
}

Since there are three possible measurement choices (i.e., $m=3$), models M1, M2, and M3 have $p=27$, $30$, and $33$ free parameters, respectively.
We also consider a “naive” (nonquantum) model M0, where each observation in $\mathbb{D}$ is treated independently under a multinomial distribution ($p=54$); using a fully overparametrized model serves as a baseline for worst-case performance.

First, we analyze state assemblages arising from isotropic states $\rho_\nu = \nu\vert \Phi^{+}\rangle\langle \Phi^{+}\vert + (1-\nu)\mathbb{I}/4$ with moderate noise $\nu=0.8$ and mean detector efficiency of $\epsilon=0.7$.
We simulate different detector bias scenarios, both without setting-dependent effects ($s=0$) and with small variance in $\epsilon_x$ ($s=0.03$).
Fig.~\ref{fig:werner_results} shows that under idealized conditions ($s=0$, $\eta \ll 1$) the simplest model performs well.
As both detector bias and variance increase, M2 is only preferable in very low bias conditions ($\eta<0.01$), whereas M3 generally minimizes information loss with respect to the ground truth.

\bgroup
\def\arraystretch{1.5}
\begin{table}
\begin{tabular}{cccccc}
 & $\Delta\mathrm{AIC}_\mathrm{M1}$ & & $\Delta\mathrm{AIC}_\mathrm{M2}$ & & $\Delta\mathrm{AIC}_\mathrm{M3}$ \\ \cline{2-6} 
\multicolumn{1}{c|}{Low bias} & 38.73 & \ \ \ & 13.61 & \ \ \ & 0 \\
\multicolumn{1}{c|}{High bias} & 885.59 & \ \ \ & 890.38 & \ \ \ & 0 
\end{tabular}
\caption{Experimental AIC differences for assemblages reconstructed from maximally entangled states $\ket{\Phi^+}$, under low ($\eta\approx 0.015$) and high ($\eta\approx 0.17$) bias conditions. 
}
\label{tab:table2}
\end{table}
\egroup

{
Our experimental validation independently corroborates these results, using a setup designed to approximate the scenario described above (SM~\cite[Note~V]{SI}).
Starting from highly pure $\ket{\Phi^+}$ states, we carry out assemblage tomography under different bias conditions.
These results are presented in Tab.~\ref{tab:table2}.
Under high bias ($\eta\approx 0.17$), M3 is the only model able to account for experimental imperfections.
Even in very low bias conditions ($\eta\approx 0.015$), M3 retains strong statistical support, aligning with our theoretical predictions.
}

\begin{figure}
  \centering
  \includegraphics[width=8.6cm]{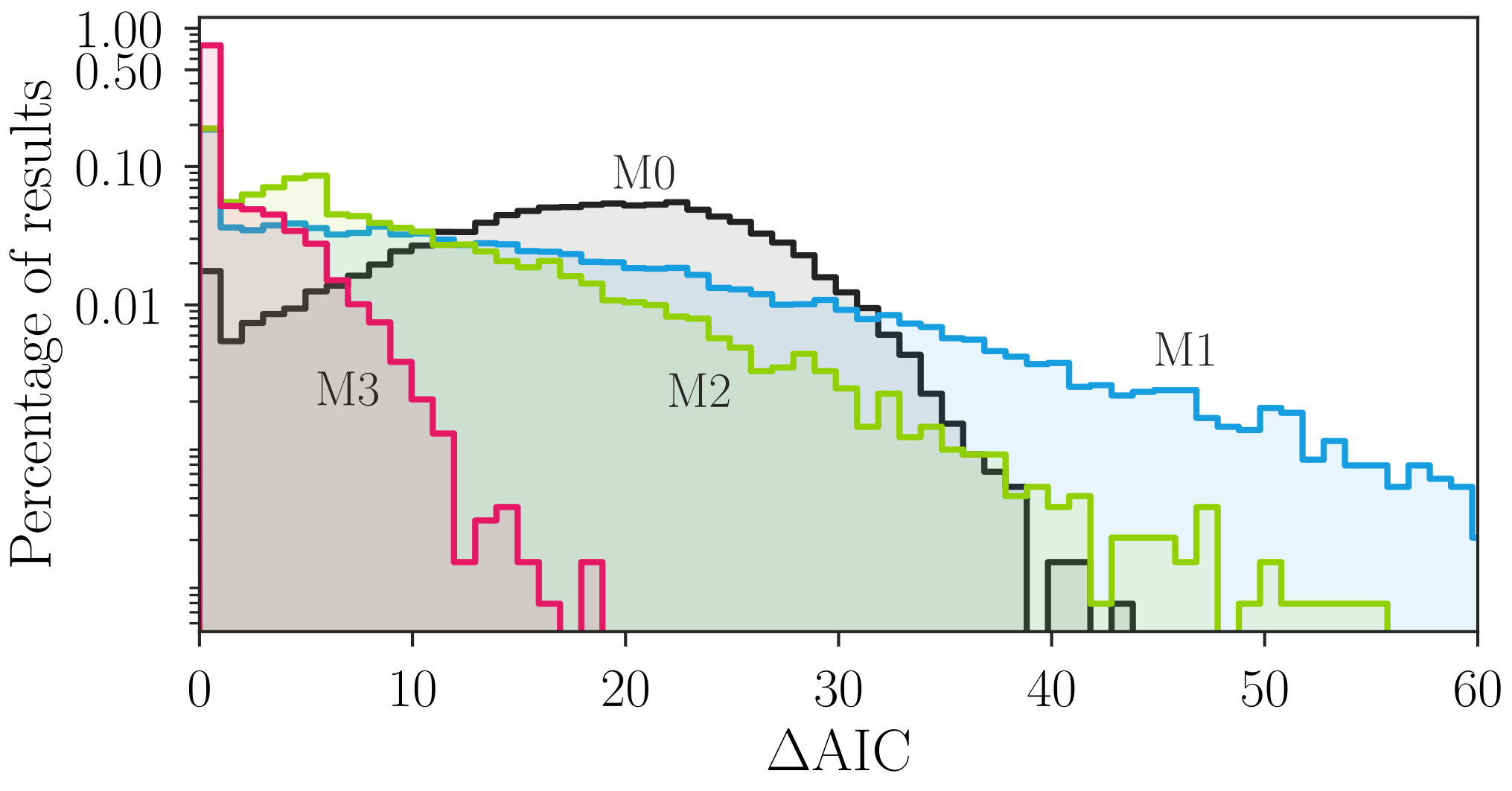}
  \caption{
  Assemblage tomography results for 15,000 randomly distributed two-qubit states with an average detection efficiency of $\epsilon = 0.7$, variance $s = 0.03$, and bias magnitude $\eta = 0.2$; the ``effective mean efficiencies" are $0.76$ and $0.7$.
  The histogram shows the distribution of $\Delta\mathrm{AIC}$ values. Model M3 is preferred for 75\% of all the simulated states.
  }
  \label{fig:random_results}
\end{figure}
\blk
As a second test case, we simulate random density matrices of two-qubit states $\rho = {AA^\dagger}/{\text{Tr}\left[AA^\dagger\right]}$, where $A$ is sampled from the Ginibre ensemble \cite{zyczkowski2011GeneratingRandomDensity}.
Fig.~\ref{fig:random_results} aggregates the results from 15,000 different states under moderate loss conditions ($s = 0.03$, $\eta = 0.2$), showing that M3 remains the preferred model even for general states.
We also include small errors in measurements and dark counts in the generated data (see SM~\cite[Note~III]{SI}).

\begin{figure}
  \centering
  \includegraphics[width=8.6cm]{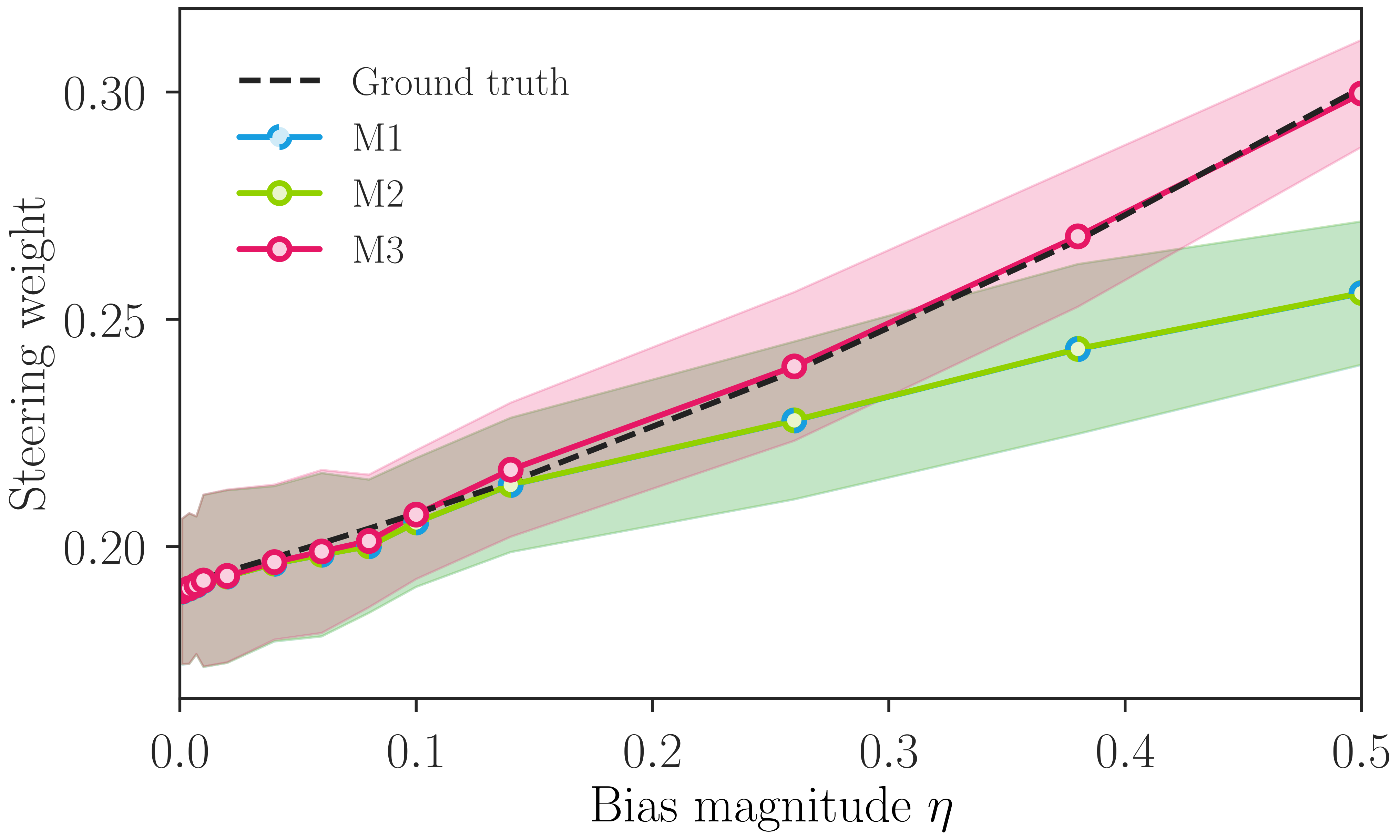}
  \caption{
  Steering weight for the numerical results of Fig.~\ref{fig:werner_results} with $\epsilon=0.7$ and $s = 0.03$ compared to a theoretical target assemblage.
  Reconstructions under models without dependence on measurement outcomes (M1 and M2) lead to a systematic underestimation of the steering weight. Uncertainties in the steering weight (shaded region) correspond to $\pm 1$ standard deviation for 500 Monte Carlo simulations.
  }
  \label{fig:steering}
\end{figure}

\parTitle{Error-tolerant quantification of steering}As assemblages play a crucial role in demonstrating quantum steering, we study how models M1--M3 may result in significantly different values of steering measures for the same dataset.
Quantum steering is formally defined as the ability to prepare assemblages that could not be produced by an appropriate averaging over an ensemble of \textit{local hidden states} (LHSs) on Bob's side, 
that is, ruling out the possibility that $\forall x, a$, $\sigma_{a|x} = \sum_\lambda p_\lambda p(a|x,\lambda)\rho_\lambda$~\cite{jones2007EntanglementEinsteinPodolskyRosenCorrelations}.
The \emph{steering weight} is a useful metric for quantifying the degree to which such a decomposition is ruled out.
Following \cite{skrzypczyk2014QuantifyingEinsteinPodolskyRosenSteering}, any assemblage $\{\sigma_{a|x}\}$ can be decomposed as a mixture of a steerable assemblage $\{\gamma^{\rm S}_{a|x}\}_{a,x}$ and an assemblage that admits an LHS $\{\sigma^{\rm LHS}_{a|x}\}_{a,x}$,
\begin{equation}
\label{eq:sw}
  \sigma_{a|x} = p \gamma^{\rm S}_{a|x} +(1-p)\sigma^{\rm LHS}_{a|x}.
\end{equation}
The steering weight of $\{\sigma_{a|x}\}$
represents the minimum value of $p$ such that a decomposition in Eq.~\eqref{eq:sw} exists, and can be computed using a single instance of a semidefinite program~\cite{skrzypczyk2014QuantifyingEinsteinPodolskyRosenSteering,cavalcanti2016QuantumSteeringReview}.
Starting from the isotropic states with $\nu=0.8$ used in Fig.~\ref{fig:werner_results}(b), we evaluate the associated steering weights of the reconstructed assemblages and compare them to the theoretical value of the ``true" assemblage.
In Fig.~\ref{fig:steering}, we see that reconstructions based on model M3 provide a close-to-unbiased estimation of the steering weight, whereas simpler loss models deviate from the true value as the imbalance in detection outcomes increases.
{ This result also holds true for the experimental data: for assemblages generated from maximally entangled $\ket{\Phi^+}$ states with $\eta\approx 0.17$, the estimated steering weight increases from $0.2904$ (M1 and M2) to $0.2972$, according to M3.}

\parTitle{Concluding remarks}As quantum steering attracts interest for its applications in quantum information tasks, providing tools for the tomography of quantum state assemblages becomes crucial. 
{Here, we introduced a robust methodology for estimating quantum assemblages designed to handle the most general loss scenarios, offering flexibility in model selection.}
Using the Akaike information criterion, we demonstrated that a general loss model is generally preferred in balancing faithfulness and complexity across a range of experimentally relevant values of loss parameters.
{However, one may use a simpler model if it is good enough or combine different models by weighting them according to their AIC values~\cite{enk2013WhenQuantumTomography}.}
Fidelity-based benchmarks, which require a known ground truth, may not effectively differentiate good and poor models, potentially leading to misinterpretations of results like the steering weight of an assemblage.
Our results suggest that even slight amounts of detector bias can largely impact the tomographic accuracy.
Therefore, while techniques like periodically swapping detector roles can reduce detector efficiency imbalances, they alone may not guarantee faithful reconstructions.
{We expect future work to build on our framework, developing benchmarking techniques for assemblage reconstruction akin to mature methods used in quantum state tomography~\cite{eisert2020QuantumCertificationBenchmarking}.}

Our results will find practical use in loophole-free demonstrations of quantum steering and other asymmetric nonlocality experiments, as it is essential to provide an accurate estimation of the detection efficiencies involved~\cite{brunner2007DetectionLoopholeAsymmetric, wittmann2012LoopholefreeEinsteinPodolsky}.
Since our approach does not depend on the dimensionality of the original state, it can be readily extended to multipartite scenarios.
{Additionally, while we focused on dichotomic measurements for simplicity, the formalism is not limited to such cases and could be used for steering in higher dimensions~\cite{zeng2022OnewayEinsteinPodolskyRosenSteering, srivastav2022QuickQuantumSteering}.}
Generally, for a POVM $\{E_{a|x}\}_a$ with $a\in \{0, 1, \ldots, n_A\}$ conclusive outcomes, one needs at most $n_A-1$ independent $\gamma_x$ terms to describe potential biases for the measurement corresponding to $x$.

\begin{acknowledgments}\emph{Acknowledgments.---}This work was supported by the Australian Research Council Centre of Excellence CE170100012 and the National Natural Science Foundation of China (No.~12288201). L.V.-A., A.P., and D.J.J.~acknowledge support from the Australian Government Research Training Program (RTP). This material is based upon work supported by the Air Force Office of Scientific Research under award number FA2386-23-1-4086.  We acknowledge the support of the Griffith University eResearch Service \& Specialised Platforms Team and the use of the High-Performance Computing Cluster ``Gowonda'' to complete this research.
The numerical conic optimization was carried out using \href{https://yalmip.github.io/}{YALMIP} \cite{yalmip} and \href{https://www.mosek.com/documentation/}{MOSEK} \cite{mosek}.
\end{acknowledgments}

%

\clearpage
\pagebreak
\widetext
\begin{center}
\textbf{\large Supplemental Material}
\end{center}

\setcounter{equation}{0}
\setcounter{figure}{0}
\setcounter{table}{0}
\setcounter{page}{1}
\makeatletter
\renewcommand{\theequation}{S\arabic{equation}}
\renewcommand{\thefigure}{S\arabic{figure}}
\renewcommand{\bibnumfmt}[1]{[S#1]}
\renewcommand{\citenumfont}[1]{S#1}

\setcounter{secnumdepth}{2}

\section{Tomography algorithm for model M3}\label{app:alg3}

\begin{figure}[ht]
	\centering
	\includegraphics[width=0.8\linewidth]{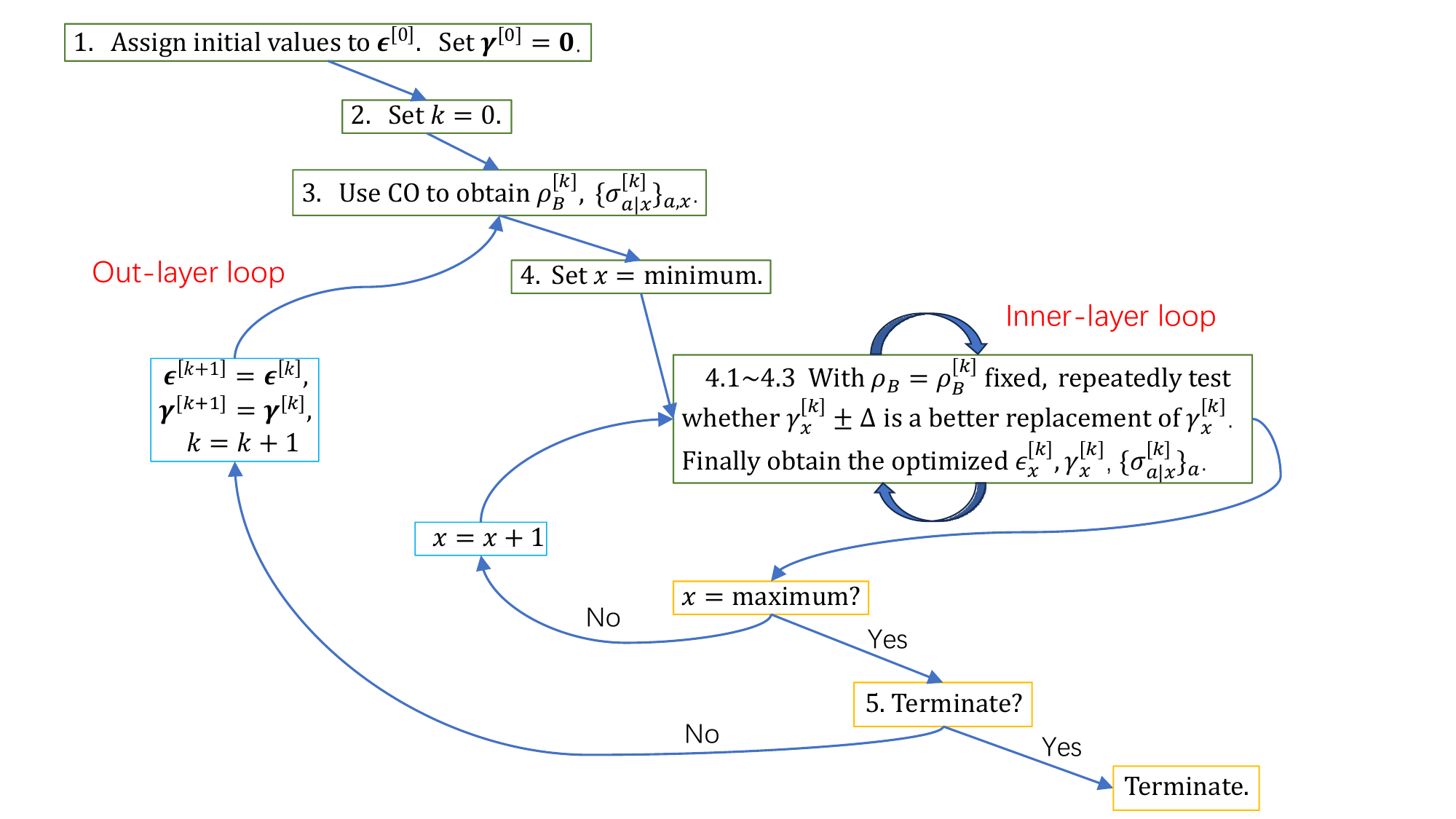}
	\caption{ Procedures of algorithm in model M3.}\label{figA2}
\end{figure}

We continue analyzing the optimization model involving Eq.~(10). Although $-\mathcal{L}$ (and, consequently, $-\ln[\mathcal{L}]$) is convex, this optimization problem is still difficult to solve because the constraint~(5) is nonlinear (in fact, nonconvex). Recall that the algorithm for M2 in Tab.~I, as a mathematically easier version than the one for M3, was solved based on conic optimization (CO) in Ref.~\cite{pepper2024ScalableMultipartySteeringz}. It thus might be helpful to analyze how far away our problem is from CO problems. We notice three facts: (i) given the values of $\boldsymbol{\gamma}$ and $\boldsymbol{\epsilon}$, we can solve $\max_{\rho_B,\{\sigma_{a|x}\}_{a,x}}\ln\left[\bar{\mathcal{L}}\right]$ using CO (here bar means all the constraints are satisfied); (ii) for each $x$, $\gamma_x$ is a scalar with only two possible updating directions; (iii) for any $x_0$, given the values of $\gamma_{x_0}$ and $\rho_{\rm B}$, we can solve $\max_{\epsilon_{x_0},\{\sigma_{a|x_0}\}_a}\ln\left[\bar{\mathcal{L}}\right]$ using CO. These motivate us to design a two-layer iterative algorithm based on CO, as shown in Fig.~\ref{figA2}: In each round of the out-layer loop, we optimize Bob's state and the assemblage, given values of the biases and efficiencies. Then with the optimization result of Bob's state in hand, for each $x$, we run an inner-layer loop independently to further optimize the bias parameter $\gamma_x$, by changing it slightly larger or smaller and retain the result with a larger cost function value. After the inner-layer loop finishes for all $x$, the optimized biases are again fed to the out-layer loop to rerun until finally the algorithm terminates. The two layers take turns to optimize part of the unknown parameters, and we thus call it a see-saw algorithm. The specific algorithm procedures of M3 are listed as follows.

\textit{Algorithm for model M3.} 

1. We employ a proper algorithm, e.g., the one used for model M2 in Tab.~I, to obtain an initial searching point to save time, and assign the obtained values of the efficiency to $\boldsymbol{\epsilon}^{[0]}$ and set $\boldsymbol{\gamma}^{[0]}=\boldsymbol{0}$.

2. Out-layer loop starts: Set $k=0$ and $\bar{\mathcal{L}}^{[-1]}=0$.

3. Out-layer optimization: Given $\{\epsilon_x^{[k]}\}_x$ and $\boldsymbol{\gamma}^{[k]}$, use CO to solve $\max_{\rho_B,\{\sigma_{a|x}\}_{a,x}}\ln\left[\bar{\mathcal{L}}\right]$ to obtain values $\rho_B^{[k]}$, $\{\sigma_{a|x}^{[k]}\}_{a,x}$ and $\ln\left[\bar{\mathcal{L}}^{[k]}\right]$.

4. Set $x=x_{\min}$ and $\bar{\mathcal{L}}^{[k]}=0$.

4.1 Inner-layer loop starts: set $j=1$. Assign the initial values as $\boldsymbol{\gamma}^{[k,1]}=\boldsymbol{\gamma}^{[k]}$, and assign $\bar{\mathcal{L}}^{[k,1]}=\bar{\mathcal{L}}^{[k,0]}=\bar{\mathcal{L}}^{[k]}$.

4.2 Inner-layer optimization: Given $\gamma_x^{[k,j]}$, calculate two tentative updates $\gamma(\pm)=\gamma_x^{[k,j]}\pm \Delta$ where $\Delta$ is determined according to $\ln\left[\bar{\mathcal{L}}^{[k,j]}\right]-\ln\left[\bar{\mathcal{L}}^{[k,j-1]}\right]$ and other necessary history information. Solve $\max_{\epsilon_x,\{\sigma_{a|x}\}_a}\ln\left[\bar{\mathcal{L}}\right]$ s.t. $\rho_B=\rho_B^{[k]}$ and $\gamma_x=\gamma(\pm)$, respectively, to obtain $\ln\left[\bar{\mathcal{L}}(\pm)\right]$. Let $\ln\left[\bar{\mathcal{L}}^{[k,j+1]}\right]=\max\{\ln\left[\bar{\mathcal{L}}(+)\right],\ln\left[\bar{\mathcal{L}}(-)\right],\ln\left[\bar{\mathcal{L}}^{[k,j]}\right]\}$ and update the corresponding $\gamma_x^{[k,j+1]}$, $\epsilon_x^{[k,j+1]}$ and $\{\sigma_{a|x}^{[k,j+1]}\}_a$.

4.3 If $\ln\left[\bar{\mathcal{L}}^{[k,j+1]}\right]-\ln\left[\bar{\mathcal{L}}^{[k,j]}\right]$ is larger than a given threshold, set $j=j+1$ and go to Step 4.2. Otherwise, set $\ln\left[\bar{\mathcal{L}}^{[k]}\right]=\ln\left[\bar{\mathcal{L}}^{[k,j+1]}\right]$, $\epsilon_x^{[k]}=\epsilon_x^{[k,j+1]}$, $\gamma_x^{[k]}=\gamma_x^{[k,j+1]}$, $\{\sigma_{a|x}^{[k]}\}_a=\{\sigma_{a|x}^{[k,j+1]}\}_a$ and $x=x+1$, and go to Step 4.1, unless $x$ has already reached its maximum value $m$, in which case go to Step 5.

5. If $\ln\left[\bar{\mathcal{L}}^{[k]}\right]-\ln\left[\bar{\mathcal{L}}^{[k-1]}\right]$ is larger than a given threshold, set $\boldsymbol{\epsilon}^{[k+1]}=\boldsymbol{\epsilon}^{[k]}$, $\boldsymbol{\gamma}^{[k+1]}=\boldsymbol{\gamma}^{[k]}$, $k=k+1$ and go to Step 3. Otherwise, terminate the algorithm and output the final estimation results as $\hat \rho_B={\hat\rho_B}^{[k]}$, $\hat{\boldsymbol\epsilon}=\boldsymbol{\epsilon}^{[k]}$, $\hat{\boldsymbol\gamma}=\boldsymbol{\gamma}^{[k]}$ and $\{\hat\sigma_{a|x}\}_{a,x}=\{\sigma_{a|x}^{[k]}\}_{a,x}$.

\subsection{Convergence analysis}
Given that $\sigma_{a|x}\leq \mathbb{I}_{\rm A}$, the cost function value in Eq.~(10) is inherently upper bounded. Neither of our two-layer loops decreases the value of $\ln\left[\bar{\mathcal{L}}\right]$, suggesting that the algorithm will always converge to a local or global optimal solution. Additionally, in our simulation, the algorithm in M3 initiates its search from the solution obtained by the algorithm in M2. Therefore, the final cost function value achieved in M3 will not be smaller than that obtained by M2.

\section{Assemblage fidelity}\label{app:fidelity}

Despite its shortcomings~\cite{peters2004MixedstateSensitivitySeveralz}, fidelity remains the most widely used metric for judging the quality of tomographic estimates.
Thus, we also use the quantity of \textit{assemblage fidelity} as defined by Ref.~\cite{nery2020DistillationQuantumSteeringz}, stressing that direct fidelity comparisons with the ground truth are only possible in theoretical studies.
For two assemblages $\{\sigma_{a\vert x}\}_{a,x}$ and $\{\xi_{a\vert x}\}_{a,x}$ with the same number of inputs $x$ and outputs $a$, the assemblage fidelity between them is defined as 
\begin{equation}
\label{eq:assfid}
  \mathcal{F}_A\left(\{\sigma_{a\vert x}\}_{a,x}, \{\xi_{a\vert x}\}_{a,x}\right) = \min_{x}\sum_{a} \mathcal{F}\left(\sigma_{a\vert x}, \xi_{a\vert x}\right),
\end{equation}
with $\mathcal{F}(\sigma,\xi) = \text{Tr}\left[\sqrt{\sqrt{\sigma}\xi\sqrt{\sigma}}\right]$ being the square root fidelity between states $\sigma$ and $\xi$.

\begin{figure}[ht]
  \centering
  \includegraphics[width=8.6cm]{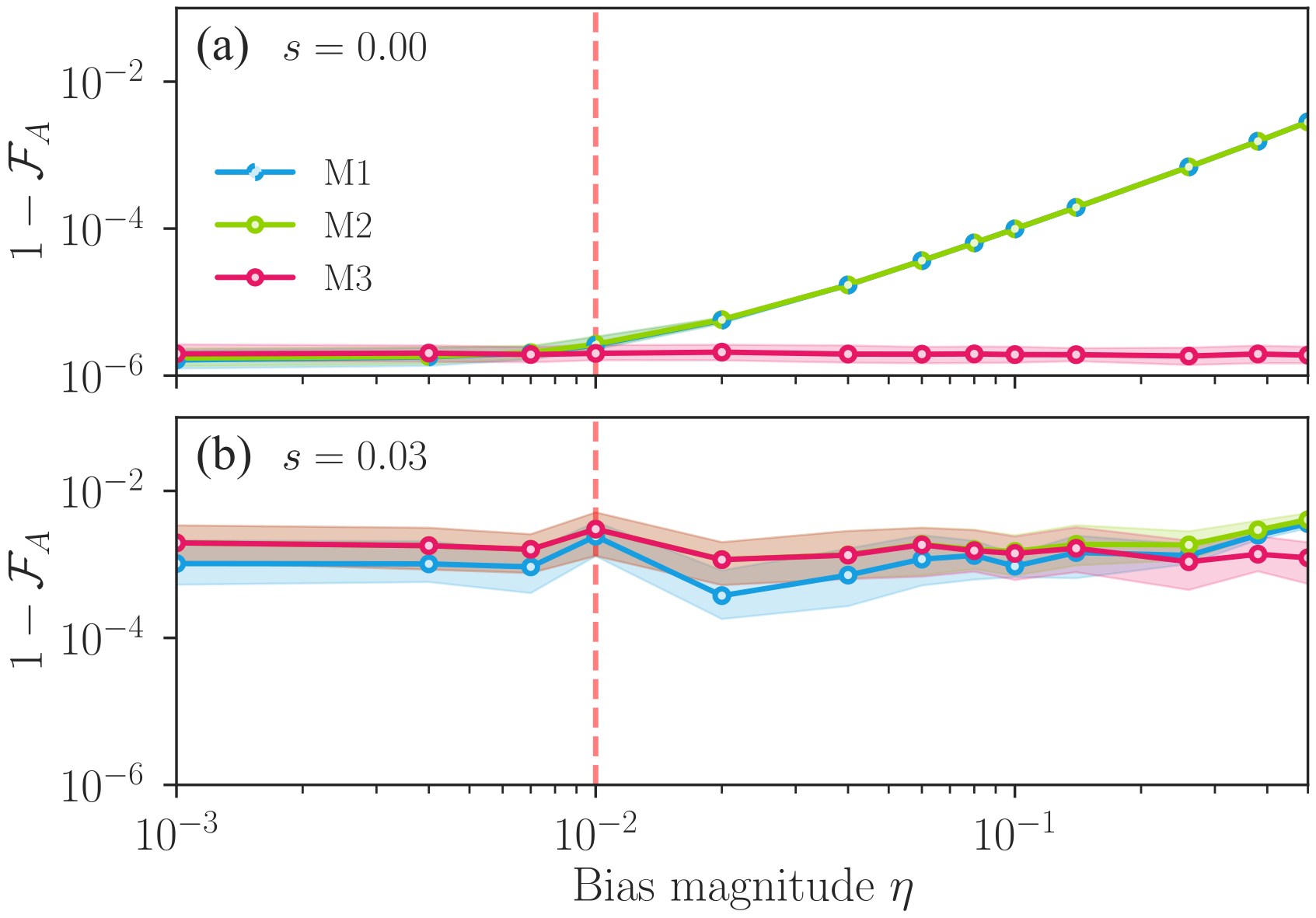}
  \caption{
  Fidelity benchmarks for the numerical simulations in Fig.~1 in the main text.
  Assemblage infidelity $1-\mathcal{F}_A$ between the estimated assemblage and an ``ideal'' target assemblage, for zero (a) and nonzero (b) setting-dependent effects.
  The vertical dashed line marks the 1\% detector bias magnitude threshold.
  }
  \label{fig:app_fidelity_results}
\end{figure}

\begin{figure}
  \centering
  \includegraphics[width=8.6cm]{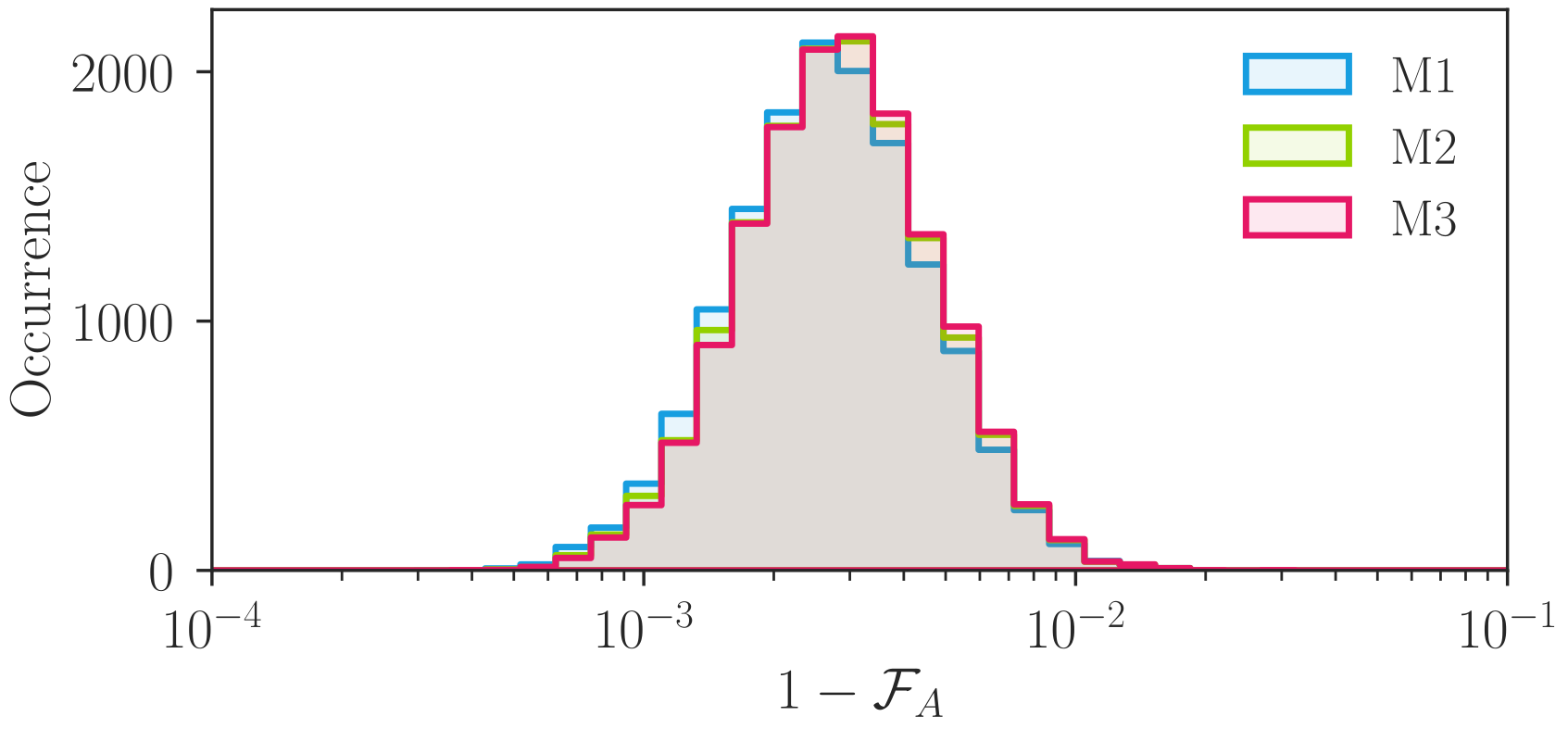}
  \caption{Fidelity benchmark for the numerical simulations in Fig.~1 in the main text.
  }
  \label{fig:app_random_results}
\end{figure}

We now reproduce the results presented in Figs.~1 and 2 in the main text, but using fidelity as the figure of merit.
In the case of Werner states (Fig.~\ref{fig:app_fidelity_results}), assemblage fidelity is less sensitive to bias when setting-dependent effects are present ($s>0$), but it broadly correlates with the $\Delta\mathrm{AIC}$ trends shown in Fig.~(1).
However, in the case of random two-qubit states (Fig.~\ref{fig:app_random_results}), assemblage fidelity struggles to distinguish between models despite high ($\mathcal{F}_A>0.99$) assemblage fidelity values across the board.

\blk

At this point, we remark that the assemblage fidelity purely aims to determine the overlap between some target ``ideal" assemblage (which is never accessible in an experiment) and a reconstruction, similar to other metrics that quantify notions of distance between two ensembles of quantum states~\cite{schumacher1996SendingEntanglementNoisyz, oreshkov2009DistinguishabilityMeasuresEnsemblesz}.
However, fidelity is known to be a particularly poor indicator of closeness between mixed quantum states.
We illustrate how this drawback might affect the notion of closeness between two assemblages that incorporate null outcomes.

First, let $\boldsymbol{\Sigma}_{A|X}(\nu)$ represent an assemblage prepared by Alice after performing measurements on the two-qubit isotropic state $\rho_\nu$.
The measurements Alice performs correspond to ``lossy" versions of the three Pauli observables $Z$, $X$, and $Y$, where her effective detection efficiency $\epsilon$ is taken to be a constant, independent from her choice of measurement and outcome; $\epsilon_x = \epsilon$ and $\gamma_x = 0~\forall x$.
In other words, we focus on the simplest scenario that corresponds to model M1.
Each element of the assemblage will then depend on $\epsilon$ and $\nu$, and will have the form:

\begin{align}
  \sigma_{+|x} &= \frac{\epsilon(\nu+1)}{4}\ketbra{x} + \frac{\epsilon(1-\nu)}{4}\ketbra{x_\perp}, \\
  \sigma_{-|x} &= \frac{\epsilon(\nu-1)}{4}\ketbra{x}- \frac{\epsilon(\nu+1)}{4}\ketbra{x_\perp}, \\
  \sigma_{\emptyset|x} &= -\frac{\epsilon\nu}{2}\ketbra{x} + \frac{\epsilon\nu}{2}\ketbra{x_\perp} + \mathbb{I}/2,
\end{align}
where $\ket{x}$ (resp.~$\ket{x_\perp}$) denotes the positive (negative) eigenvector of the Pauli operator associated with that particular measurement.
For example, in the case of $Z$, $\ket{x} = \ket{0}$ and $\ket{x_\perp} = \ket{1}$ and similarly for $X$ and $Y$. 

\begin{figure}
  \centering
  \includegraphics[width=12.9cm]{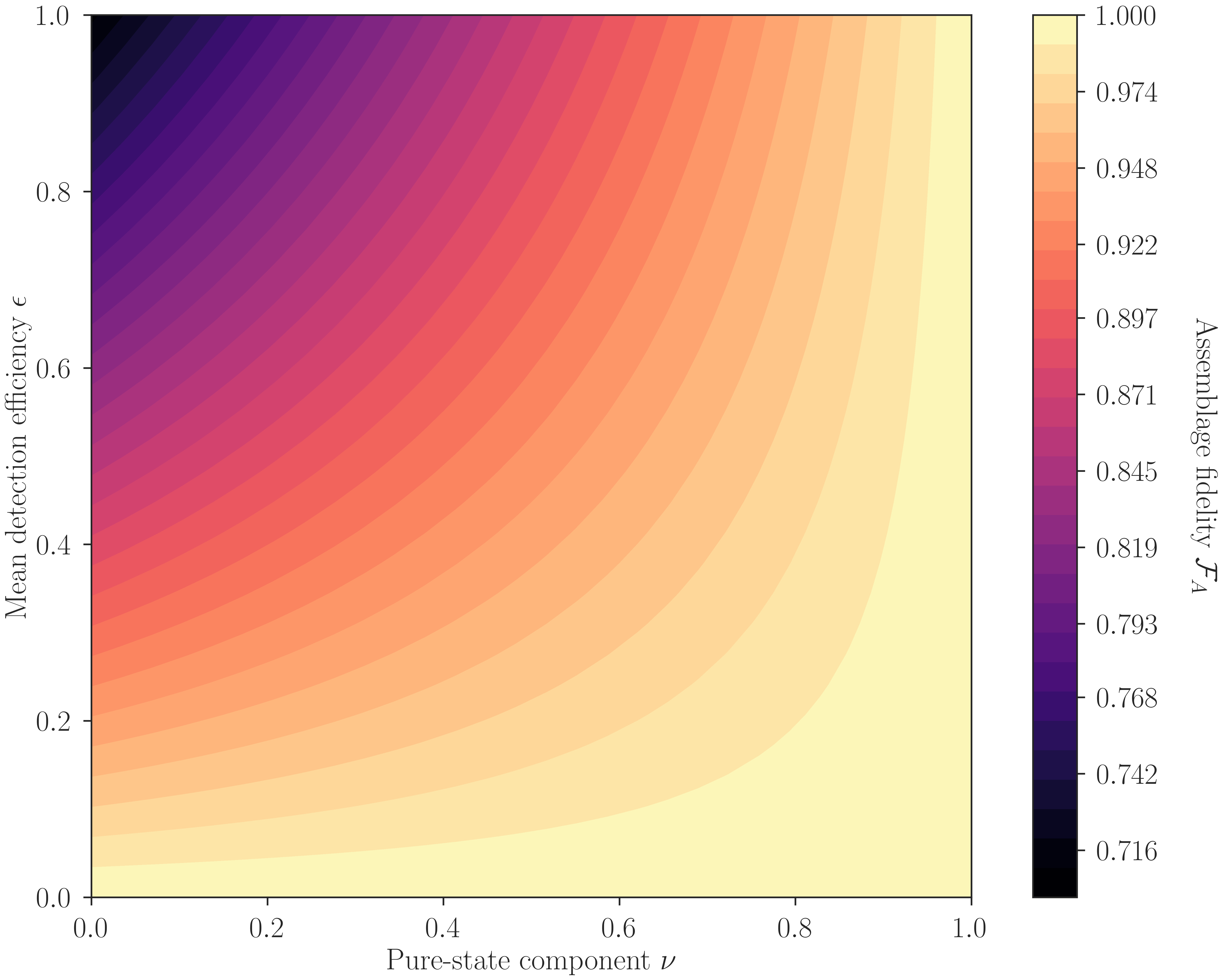}
  \caption{Sensitivity of assemblage fidelity in distinguishing an assemblage prepared from the maximally entangled state $\vert\Phi^{+}\rangle$ and an assemblage prepared from the same state with added white noise $\rho_\nu = \nu\ketbra{\Phi^+} + (1-\nu)\mathbb{I}/4$.
  As the mean detection efficiency decreases, assemblage fidelity becomes less effective at differentiating between the two.
  }
  \label{fig:app_fidelity}
\end{figure}

We are now interested in computing the assemblage fidelity, as defined in the main text, between $\boldsymbol{\Sigma}_{A|X}(\nu \leq 1)$ (generated from a partially mixed state) and $\boldsymbol{\Xi}_{A|X}(\nu=1)$, i.e., an assemblage obtained from measurements performed on a pure, maximally entangled state.
As $\epsilon$ decreases, the impact of states associated with a null outcome becomes increasingly dominant in the calculation of the fidelity.
We illustrate this in Fig.~\ref{fig:app_fidelity}.
In the limit of low efficiency, a large enough assemblage fidelity (say, $\mathcal{F}_A>0.99$) could originate from a very pure maximally entangled state shared between Alice and Bob, or from a completely mixed one.

\section{Simulation details}\label{app:simulation}
In our numerical simulations, we encode qubits in the polarization mode of single photons emitted by a source following a Poissonian distribution. The measurements performed by Bob involve projections onto the tomographically complete set $\{H, V, D, A, R, L\}$, with each measurement station simulating a quarter-wave plate, a half-wave plate, and a polarizing beam-splitter. We consider two independent single-photon counters at each output of the beam-splitter, accounting for $+$ and $-$ outcomes.

Alice's detection is modeled following the parametrization described in the main text.
As we assume Bob to be honest with his measurements, we only consider the measurement rounds where his photon is not lost. Therefore, his measurements can always be regarded as lossless.

All the simulations are then performed in the following way:
\begin{enumerate}
  \item Generate $N$ quantum states according to a prior distribution. Here, we considered $N=10^5$.
  \item Simulate lossy (lossless) measurements for Alice (Bob), generating a dataset of joint outcomes. We set $\epsilon=0.7$ as Alice's mean efficiency.
  \item Estimate the maximum likelihood in Eq.~(10) and obtain an estimate of $\{\sigma_{a\vert x}\}_{a,x}$ according to each loss model.
  \item Calculate the AIC values for each model and compute the assemblage fidelity between each reconstruction and the ``true" assemblage.
\end{enumerate}

For assemblages obtained from random states, we also consider small errors in the measurements performed by Bob.
Including nonideal measurements provides a better representation of real experimental conditions and has been studied in the context of steering tasks~\cite{bennet2012ArbitrarilyLossTolerantEinsteinPodolskyRosenz, tavakoli2024QuantumSteeringImprecisez}. 
We incorporate both statistical and systematic errors in the simulated dataset prepared for Bob.
Since our experiments are based on quantum optics, we consider intrinsic fluctuations in the probabilistic generation and detection of single photons, which we assume follow a Poisson distribution.
Current single-photon detectors also suffer from dark counts. Here we consider mean dark count values of $\approx 10^2$ per joint outcome.

Systematic errors in photonic systems might stem from manufacturing imperfections in the optical and mechanical elements used (like measurement wave plates being controlled by motorized mounts).
We choose conservative magnitudes of these errors: we assume a wave plate retardation error of $|\lambda/120|$, as well as motorized rotation stage precision of $0.08^\circ$.
Each trial samples independently from a Poisson distribution for the statistical errors, and from a normal distribution for the systematic ones.

\section{Effects of sample size on the AIC}\label{app:sample_size}

When dealing with small sample sizes, the AIC may exhibit a bias towards more complex models, potentially leading to overfitting. To address this, one can use the small-sample-size-corrected version of AIC, known as AICc, which adjusts for the reduced data availability by incorporating a second-order penalty term on the number of model parameters $p$~\cite{hurvich1989RegressionTimeSeriesz}.
Specifically, the AICc is defined as:
\begin{equation}
  \mathrm{AICc} = \mathrm{AIC} + \frac{2p^2 + 2p}{n-p-1},
\end{equation}
where $n$ is the size of the dataset $\mathbb{D}$. 

\begin{figure}[ht]
  \centering
  \includegraphics[width=8.6cm]{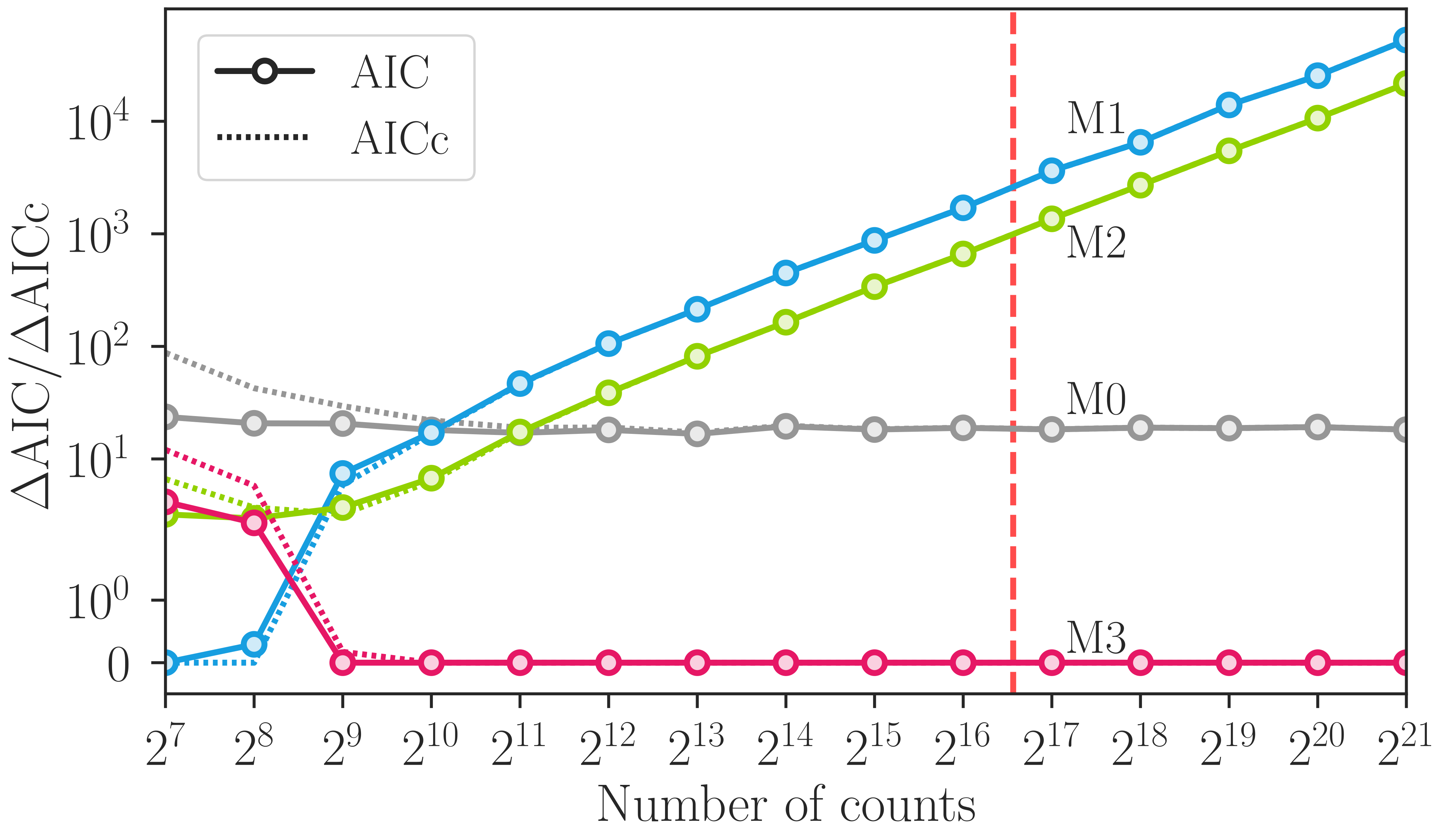}
  \caption{Comparison between the $\Delta\mathrm{AIC}$ (solid lines with markers) and its small sample size corrected version $\Delta\mathrm{AICc}$ (dotted lines) as the sample size increases.
  Results are based on Fig.~(1), with $\epsilon=0.7$, $s=0.03$, and $\eta=0.2$.
  As the number of experimental counts grows, both criteria converge, with model M3 emerging as the one minimizing information loss.
  The dashed vertical line represents the number of counts used in our simulations ($N=10^5$).
  }
  \label{fig:samplesize}
\end{figure}

We show the effect of sample size on our simulations in Fig.~\ref{fig:samplesize}, based on the results presented in Fig.~(1).
For larger enough datasets, the AICc converges to the AIC, with reconstructions following the general model M3 becoming increasingly supported by the data.
Given the large sample size in our simulations ($N = 10^5$, vertical dashed line), the impact of the sample size correction can be considered negligible.

\pagebreak

{
\section{Experimental details and data}\label{app:experiment}
The experimental data is obtained from the same source used in Ref.~\cite{joch2022CertifiedRandomnumberGenerationz}, which is designed to produce high-quality polarization-entangled states via spontaneous parametric down-conversion.
A detailed description of the source can be found in Ref.~\cite{tischler2018ConclusiveExperimentalDemonstrationz}.
The experimental states have an Uhlmann-Jozsa~\cite{jozsa1994FidelityMixedQuantumz} fidelity of $\mathcal{F} = 0.9926\pm0.0001$ with the two-qubit state $\ket{\Phi^+} =\left(\ket{HH}+\ket{VV}\right)/\sqrt{2}$, where $H$ and $V$ denote horizontal and vertical polarization, respectively.

The experimental dataset for the assemblage tomography is generated from polarization measurements performed on both qubits.
Each measurement station consists of a quarter-wave plate, a half-wave plate, and a polarizing beam splitter.
Superconducting nanowire detectors~\cite{esmaeilzadeh2017SinglephotonDetectorsCombiningz}, with efficiencies ranging from 80\% to 90\%, are placed at each output port of the beam splitter, and detection events are recorded using a time tagger.
During each measurement round, Alice and Bob independently choose to measure one of the observable $X$, $Y$, or $Z$.
Alice reports an outcome $a\in\{+,-\}$ if one of her detectors clicks, or $a=\emptyset$ if neither of them does.
Bob post-selects the rounds in which one of his detectors records an outcome, i.e., $b\in\{+,-\}$.

Since the detectors used have slightly different efficiencies, we explore two different experimental scenarios: one in which Alice's detection outcomes are inherently biased (i.e., one of her detectors is more efficient than the other) and another one in which Alice actively tries to minimize said bias by averaging out her detection efficiencies.
This efficiency averaging operation is done by periodically swapping the role of the detectors associated with the positive and negative outcomes throughout the experiment, a strategy commonly employed in photonic experiments.
To minimize the influence of Bob's own detector imbalance in the results (and because he is considered a trusted party), we average the efficiency of his detectors in both scenarios.
We carry out assemblage tomography as per our three models and calculate the AIC differences. 
The experimental values of $\eta$ used in Tab.~II are an average derived from all the entries of the corresponding $\vec\gamma$ and $\vec\epsilon$ estimated during tomography.

We show the experimental datasets 
in Tabs.~\ref{tab:table3} and \ref{tab:table4}.
When Alice's detection efficiencies are not balanced (a high-bias scenario), the experimental results ($\Delta\mathrm{AIC}_\mathrm{M1} = 885.59$, $\Delta\mathrm{AIC}_\mathrm{M2} =890.38$, $\Delta\mathrm{AIC}_\mathrm{M3} = 0$)
give very strong statistical support for using the general loss model $\mathrm{M3}$ in the tomography task.

This result is still true even when Alice actively tries to minimize bias by averaging her efficiencies (i.e., a low-bias scenario).
The experimental results in this case ($\Delta\mathrm{AIC}_\mathrm{M1} = 38.73$, $\Delta\mathrm{AIC}_\mathrm{M2} = 13.61$, $\Delta\mathrm{AIC}_\mathrm{M3} = 0$) suggest that there is still an underlying outcome bias in the detectors.
Here too, the general loss model explains the experiment's data better than simpler models.

\bgroup
\def\arraystretch{1.5}
\begin{table*}[ht]

{\setlength{\tabcolsep}{1em}
\begin{tabular}{lrrrrrrrrrrr}
 & $a=+$ &  & $a=+$ &  & $a=-$ &  & $a=-$ &  & $a=\emptyset$ &  & $a=\emptyset$ \\
 & $b=+$ &  & $b=-$ &  & $b=+$ &  & $b=-$ &  & $b=+$ &  & $b=-$ \\ \cline{2-12} 
\multicolumn{1}{l|}{$Z\otimes Z$} & 47104 &  & 173 &  & 223 &  & 42666 &  & 37597 &  & 41940 \\
\multicolumn{1}{l|}{$Z\otimes X$} & 23735 &  & 23416 &  & 21434 &  & 21612 &  & 38672 &  & 39044 \\
\multicolumn{1}{l|}{$Z\otimes Y$} & 23377 &  & 23857 &  & 21939 &  & 21303 &  & 39285 &  & 39553 \\
\multicolumn{1}{l|}{$X\otimes Z$} & 23486 &  & 23625 &  & 21792 &  & 21459 &  & 40132 &  & 39682 \\
\multicolumn{1}{l|}{$X\otimes X$} & 47262 &  & 197 &  & 283 &  & 44146 &  & 37884 &  & 41260 \\
\multicolumn{1}{l|}{$X\otimes Y$} & 23085 &  & 24487 &  & 22683 &  & 21374 &  & 39514 &  & 39313 \\
\multicolumn{1}{l|}{$Y\otimes Z$} & 23971 &  & 23030 &  & 21429 &  & 22282 &  & 39909 &  & 40385 \\
\multicolumn{1}{l|}{$Y\otimes X$} & 23957 &  & 24183 &  & 22245 &  & 22134 &  & 40117 &  & 40054 \\
\multicolumn{1}{l|}{$Y\otimes Y$} & 205 &  & 48362 &  & 43733 &  & 288 &  & 42352 &  & 37238  
\end{tabular}}
\caption{Experimental counts for quantum assemblage reconstruction.
The efficiency of Bob's detectors is averaged by swapping the role of his detectors; the efficiency of Alice's detectors is not.
Rows indicate the measurement choice of Alice and Bob, and columns denote the joint outcomes of Alice ($a\in\{+,-,\emptyset\}$) and Bob ($b\in\{+,-\}$).
}
\label{tab:table3}
\end{table*}
\egroup

\bgroup
\def\arraystretch{1.5}
\begin{table*}[ht]

{\setlength{\tabcolsep}{1em}
\begin{tabular}{lrrrrrrrrrrr}
 & $a=+$ &  & $a=+$ &  & $a=-$ &  & $a=-$ &  & $a=\emptyset$ &  & $a=\emptyset$ \\
 & $b=+$ &  & $b=-$ &  & $b=+$ &  & $b=-$ &  & $b=+$ &  & $b=-$ \\ \cline{2-12} 
\multicolumn{1}{l|}{$Z\otimes Z$} & 89403 &  & 418 &  & 406 &  & 89025 &  & 79559  &  & 81125  \\
\multicolumn{1}{l|}{$Z\otimes X$} & 45591 &  & 44805 &  & 45009 &  & 45305 &  & 78056  &  & 79277  \\
\multicolumn{1}{l|}{$Z\otimes Y$} & 44635 &  & 45872 &  & 45793 &  & 44532 &  & 78575  &  & 80587  \\
\multicolumn{1}{l|}{$X\otimes Z$} & 45128 &  & 45213 &  & 45240 &  & 44656 &  & 78256  &  & 81167  \\
\multicolumn{1}{l|}{$X\otimes X$} & 91380 &  & 420 &  & 463 &  & 91268 &  & 76814  &  & 79250  \\
\multicolumn{1}{l|}{$X\otimes Y$} & 44226 &  & 47113 &  & 47273 &  & 44725 &  & 77923  &  & 78761  \\
\multicolumn{1}{l|}{$Y\otimes Z$} & 46314 &  & 44516 &  & 44236 &  & 46594 &  & 79984  &  & 80684  \\
\multicolumn{1}{l|}{$Y\otimes X$} & 46273 &  & 45247 &  & 45537 &  & 46378 &  & 79856  &  & 79832  \\
\multicolumn{1}{l|}{$Y\otimes Y$} & 473 &  & 91149 &  & 91381 &  & 456 &  & 79615  &  & 78942 
\end{tabular}}
\caption{Same as Tab.~\ref{tab:table3}, with the difference that the efficiency of Alice's detection efficiency is now also averaged during the experiment.
}
\label{tab:table4}
\end{table*}
\egroup
}

\newpage

%

\end{document}